\documentclass[11pt]{article}%{report}
\usepackage{amssymb}
\usepackage{amsmath}
\usepackage{lscape}
\usepackage[dvips]{epsfig}
\def\nuc#1#2{\relax\ifmmode{}^{#1}{\protect\text{#2}}\else${}^{#1}$#2\fi}
\newcommand{\dsp}{\displaystyle}
\makeatletter
\newcommand{\l@vveden}[2]{\hbox to\textwidth{{\bf \quad #1 #2}}}

\makeatother

\begin{document}

{\bf \Large 
\noindent{
LANL Report LA-UR-00-3597, Los Alamos (2000)\\
}}

%\vspace*{0.2cm}
{\bf \Large 
\noindent{
Proc. SATIF5, July 18-21, 2000, Paris, France}}

\vspace*{2.0cm}
\begin{center}

{\bf
STUDY OF RESIDUAL PRODUCT NUCLIDE YIELDS IN 1.0 GEV PROTON-IRRADIATED 
\nuc{208}{Pb} AND 2.6 GEV PROTON-IRRADIATED \nuc{nat}{W} THIN TARGETS} 
%Study of Residual Product Nuclide Yields in 1.0 GeV Proton Irradiated \nuc{208}{Pb} and 2.6 GeV Proton Irradiated \nuc{nat}{W} Thin Targets }

\vspace{1.2cm}

{\bf Yury E.~Titarenko, Oleg V.~Shvedov, Vyacheslav F.~Batyaev, 
Valery M.~Zhivun, \\
Evgeny I.~Karpikhin, Ruslan D.~Mulambetov, Dmitry V.~Fischenko, \\ 
Svetlana V.~Kvasova}

 Institute for Theoretical and Experimental Physics,
B. Cheremushkinskaya 25, 117259 Moscow, Russia
%e-mail: Yury.Titarenko@itep.ru} \vspace{0.1cm} \\

\vspace{0.1cm}
{\bf Stepan G.~Mashnik, Richard E.~Prael, Arnold J.~Sierk}

 Los Alamos National Laboratory, Los Alamos, NM 87545, USA 

\vspace{0.1cm}
{\bf Hideshi Yasuda}

Japan Atomic Energy Research Institute, Tokai, Ibaraki, 319-1195, Japan

\vspace{1.0cm}

\end{center}

\section*{Abstract}
\noindent

113 residual product nuclide yields in a 1.0 GeV proton-irradiated 
thin monoisotopic \nuc{208}{Pb} 
sample and 107 residual product nuclide yields in a 2.6 GeV proton-irradiated 
\nuc{nat}{W} sample have 
been measured and simulated by 8 different codes. The irradiations were made 
using proton beams 
extracted from the ITEP synchrotron. The nuclide yields were 
$\gamma$-spectrometered directly using a high-resolution Ge-detector. 
The $\gamma$-spectra were processed by the GENIE-2000 code. The ITEP-developed
SIGMA code was used together with the PCNUDAT nuclear decay database to 
identify the $\gamma$-lines and 
to determine the cross-sections. The  \nuc{27}{Al}(p,x)\nuc{22}{Na}  reaction 
was used to monitor the proton flux.
The measured yields are compared with calculation
the LAHET (with ISABEL and Bertini options), CEM95, CEM2k,
CASCADE, CASCADE/INPE, INUCL, and YIELDX codes. Estimates of the 
mean deviation factor are used to 
demonstrate the predictive power of the codes. The results obtained 
may be of interest in studying the 
parameters of the Pb and W target modules of the hybrid Accelerator-Driven 
System (ADS) facilities. 

\section*{Introduction}

At present, the Pb-Bi eutectic and W are regarded as the most promising 
target materials for ADS 
facilities \cite{bib1}-\cite{bib3}. As a result,
the high-energy irradiation mode of using the materials necessitates 
additional studies of the nuclear-physics characteristics of Pb, Bi, 
and W, particularly the yields of 
residual product nuclei under proton irradiation in a broad range of energies from a few MeV to 2-3 
GeV. 
Results of such studies are extremely important when designing 
even demonstration versions of the ADS facilities. 

Undoubtedly, computational methods will play an important role when 
forming a set of 
nuclear constants for ADS facilities. Therefore, verification of the most 
extensively used simulation codes has proved to be of a high priority.

\section*{Basic definitions and computational relations}

The formalism of representing the reaction product yields 
(cross sections) in high-energy proton-irradiated thin targets 
is described in sufficient detail in \cite{bib4}. In terms of the formalism, 
the variations in 
the concentration of any two chain nuclides produced in an irradiated 
target ($N_1\xrightarrow{\lambda_1}N_2\xrightarrow{\lambda_2}$) 
may be presented to be a set of differential equations that describe 
the production and decays of the 
nuclides. By introducing a formal representation of the time functions,
$F_i$, of the form 
$F_i=\left( 1-e^{-\lambda_i\tau} \right) \frac{\dsp 1-e^{-\lambda_iKT}}
{\dsp 1-e^{-\lambda_iT}}\mbox{ ,} \quad$ (i=1, 2, and Na
or another monitor product; 
$\tau$  is the duration of a single proton pulse; $T$ is the pulse 
repetition period; $K$ is the number of pulses within the irradiation period), 
which characterize the nuclide decays 
within the irradiation time, and by expressing (similar to the relative 
measurements) the proton fluence 
size via the monitor reaction cross section, $\sigma_{st}$, we can present 
the cumulative and independent yields as:
\begin{equation}\label{1}
\sigma_1^{cum} = \frac{A_0}{\eta_1\varepsilon_1F_1N_{Na}}
\frac{N_{Al}}{N_T}\frac{F_{Na}}{\lambda_{Na}}\;\sigma_{st}  
\end{equation}
\begin{equation}
\sigma_1^{cum} = \frac{A_1}{\nu_1\eta_2\varepsilon_2F_1N_{Na}}
\frac{N_{Al}}{N_T}\frac{\lambda_2-\lambda_1}{\lambda_2}
\frac{F_{Na}}{\lambda_{Na}}\;\sigma_{st}  
\label{1a} 
\end{equation}
\begin{equation}
\sigma_2^{ind} = \left(\frac{A_2}{F_2}+\frac{A_1}{F_1}
\frac{\lambda_1}{\lambda_2}\right)
\frac{1}{\eta_2\varepsilon_2N_{Na}}
\frac{N_{Al}}{N_T}\frac{F_{Na}}{\lambda_{Na}}\;\sigma_{st} 
\label{2}  
\end{equation}
\begin{equation}
\sigma_2^{cum} = \sigma_2^{ind}+\nu_1\sigma_1^{cum} 
= \dsp \left( \frac{A_1}{F_1}+\frac{A_2}{F_2} \right)
\frac{1}{\eta_2\varepsilon_2N_{Na}}
\frac{N_{Al}}{N_T}\frac{F_{Na}}{\lambda_{Na}}
\;\sigma_{st}  \mbox{ ,} \label{3} 
\end{equation}
where $\sigma_1^{cum}$ is the cumulative cross section of the 
first nuclide; $\sigma_2^{ind}$ and 
$\sigma_2^{cum}$  are the 
independent and cumulative cross sections of the second nuclide; 
N$_{Al}$  and N$_T$  are the numbers of nuclei in the monitor 
(standard) and in experimental sample, respectively; 
$\eta_1$ and $\eta_2$ are
the $\gamma$-line yields; 
$\varepsilon_1$ and $\varepsilon_2$ are the spectrometer efficiencies 
at energies 
E$_{\gamma_1}$ and E$_{\gamma_2}$; $\nu_1$ is the branching 
ratio of the first nuclide; $\lambda_1$, $\lambda_2$, and $\lambda_{Na}$ 
are, respectively, the decay 
constants of the first and second 
nuclides and of the monitor product (\nuc{22}{Na} and/or \nuc{24}{Na}). 

The factors A$_0$, A$_1$, and A$_2$  are calculated through fitting 
the measured counting rates in the total 
absorption peaks, which correspond to energies E$_{\gamma_1}$  (the 
first nuclide) and 
E$_{\gamma_2}$  (the second nuclide), by 
exponential functions. It should be noted that formulas 
(\ref{1})--(\ref{3}) were derived on assumption that the $\gamma$-intensities 
of the two nuclides produced under irradiation are recorded up to the desired 
accuracy within an interval of time 
from the irradiation end to the moment of the ultimate detectable intensity. 
If, for some reasons, the 
factor A$_1$ cannot be found, then the factor A$_2$ will be used together 
with expression (14) from \cite{bib4} to 
determine the quantity $\sigma_2^{cum^*}$, which we called 
the supra cumulative yield:
 \begin{eqnarray}\label{sig2} 
\dsp \sigma_2^{cum^*}&=&\sigma_2 + \frac{\lambda_2}{\lambda_1-\lambda_2} 
\nu_1 \sigma_1^{cum} = \nonumber \\  
\dsp &=& \frac{A_2}{\eta_2\varepsilon_2F_2N_{Na}}
\frac{N_{Al}}{N_T}\frac{F_{Na}}{\lambda_{Na}}\sigma_{st}  
%\frac{A_2}{N_{T} \Phi \eta_2 \varepsilon_2 F_2} 
\end{eqnarray}

The resultant value of $\sigma_2^{cum^*}$  may prove to be very different 
from $\sigma_2^{cum}$. Nevertheless, the supra cumulative 
yield can be either used directly to verify the codes, or determined 
further up to $\sigma_2^{cum}$ if the needed data are 
obtained elsewhere (for example, from inverse-kinematics experiments).

\section*{Experimental techniques}
\noindent

A 10.5 mm diameter, 139.4 mg/cm$^2$ monoisotopic \nuc{208}{Pb} metal 
foil sample (97.2\% \nuc{208}{Pb}, 
1.93\% \nuc{207}{Pb}, 0.87\% \nuc{206}{Pb}, $<$0.01\% \nuc{204}{Pb}, 
$<$ 0.00105\% of chemical impurities) and a 38.1 mg/cm$^2$ \nuc{nat}{W} 
metal foil sample (99.95\% W, $<$0.05\% of chemical impurities), both of 
10.5-cm diameter, were proton-irradiated. 139.6 mg/cm$^2$ and 139.1 
mg/cm$^2$ Al foils of the same diameter were used as monitors. 
Chemical impurities of the monitor did not exceed 0.001\%.

The samples were irradiated by the external proton beam from the 
ITEP U-10 synchrotron \cite{bib4}. 
The average flux densities during irradiation of Pb and W samples 
were of 1.4$\times$10$^{10}$ p/cm$^2$ 
and 2.8$\times$10$^{10}$ p/cm$^2$, respectively.

 Our measurements were supported by extra researches aimed at 
 reducing the systematic errors in 
the experimental results. Theses researches included:
\begin{itemize}
 \item experiments to specify the neutron component in the extracted 
 proton beams,
 \item experiments to specify the \nuc{27}{Al}(p,x)\nuc{24}{Na} 
 monitor reaction cross section,
 \item studies to specify the dependence of the $\gamma$-spectrometer 
 detection efficiency on the position 
geometry of irradiated sample,
 \item studies to optimize the $\gamma$-spectrum simulation codes.
\end{itemize}

\begin{figure} %\vspace{1cm} 
\begin{center}
\includegraphics[width=10.3cm]{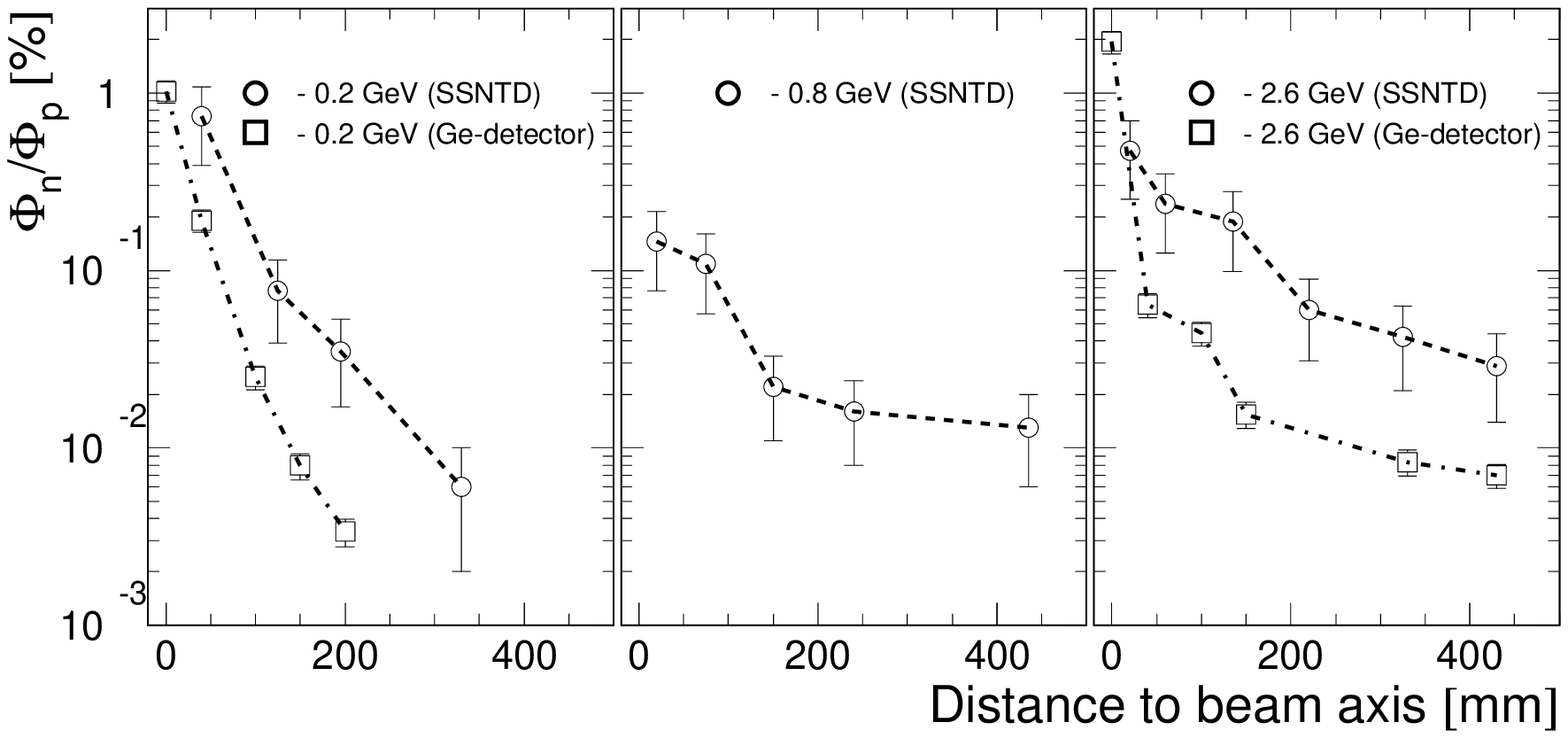}
\hfill
\includegraphics[width=4.7cm]{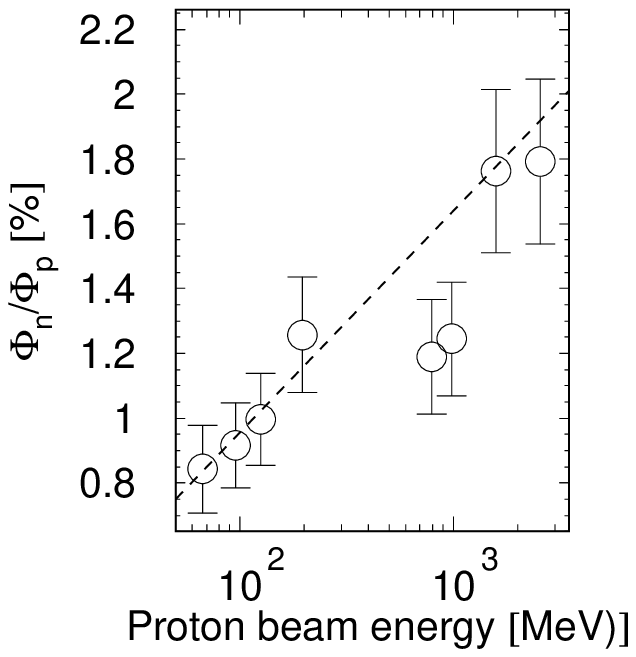}
\\
\parbox[t]{10.3cm}{\caption{The neutron backgrounds around the 
extracted proton 
beams that irradiate thin experimental samples.}\label{fig1}}
\hfill
\parbox[t]{4.7cm}{\caption{Neutron component in 
the extracted proton beams of 
different energies.}\label{fig2}}
\end{center} \end{figure}

\begin{figure} %\vspace{1cm} 
\begin{center}
\includegraphics[width=7.2cm]{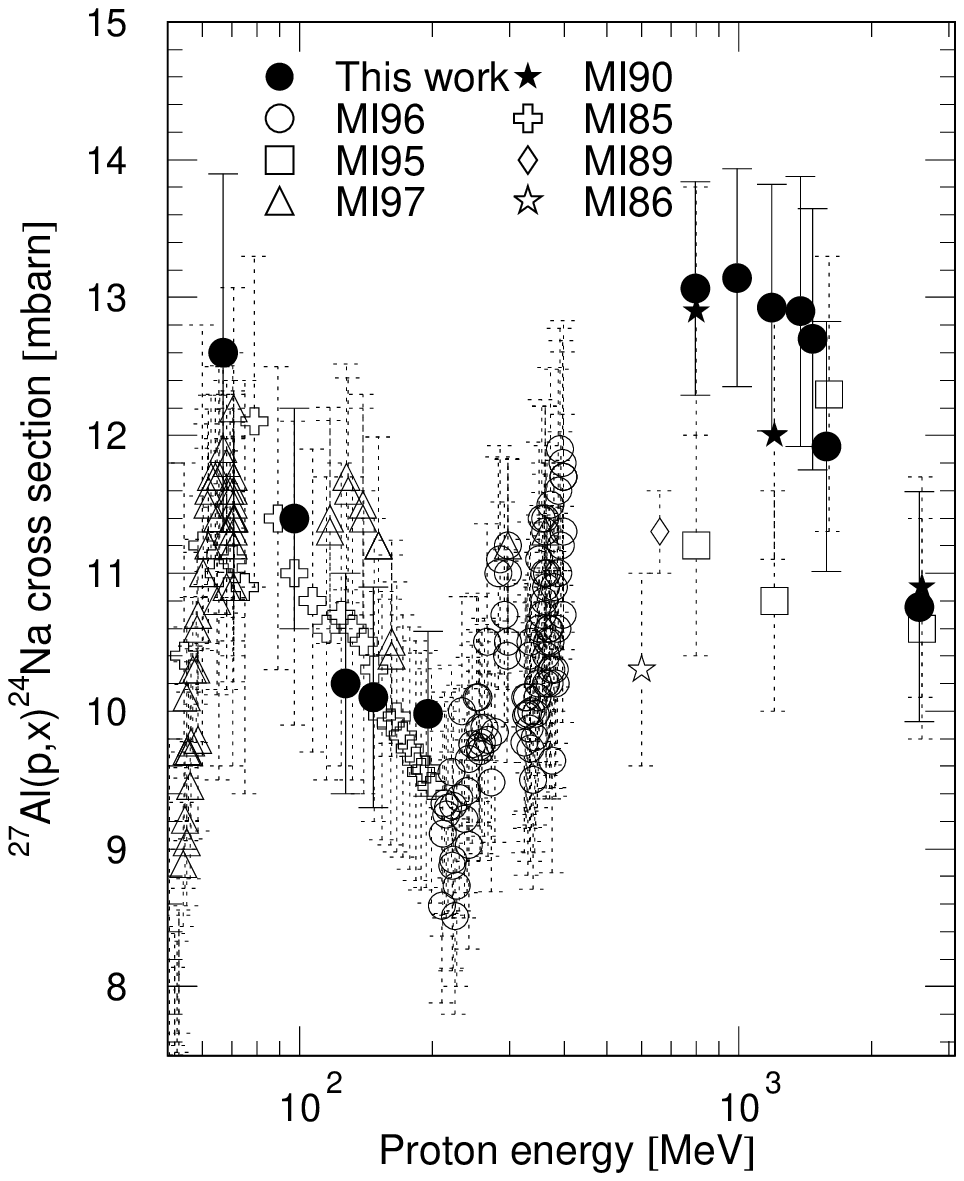}
\hfill
\includegraphics[width=7.8cm]{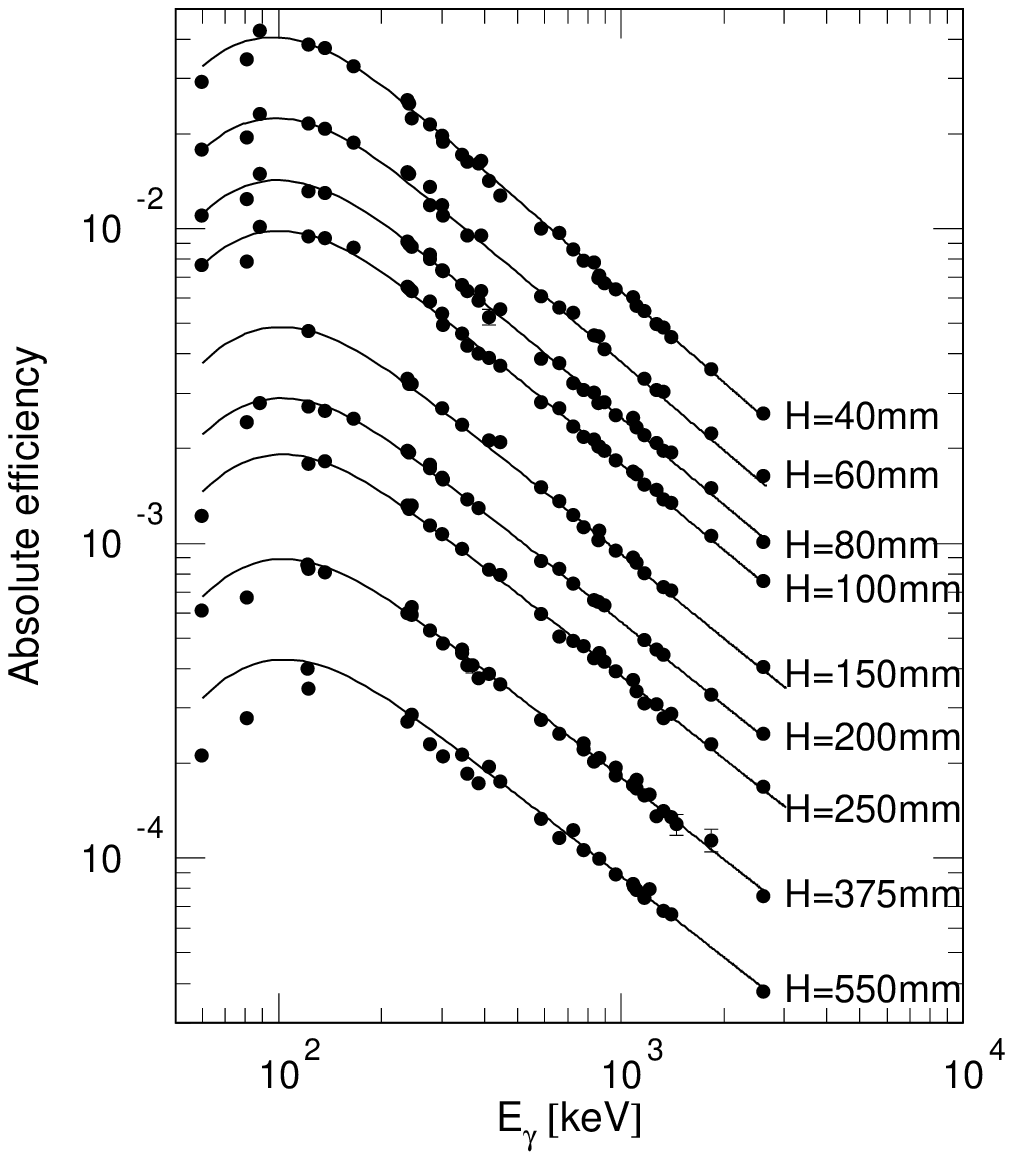}
\\
\parbox[t]{7.2cm}{\caption{The $^{27}$Al(p,x)$^{24}$Na monitor reaction cross 
sections measured in this and other 
%([5] and references therein) 
works.}\label{fig3}}
\hfill
\parbox[t]{7.8cm}{\caption{The experimental and calculated 
detection efficiency of the 
spectrometer.}\label{fig4}}
\end{center} \end{figure}

Figs. \ref{fig1} and \ref{fig2} show the results of measuring the 
neutron component in the extracted proton beams, 
i.e., the neutron-to-proton flux density ratio, $\Phi_n/\Phi_p$. 
Fig. \ref{fig3} presents the monitor reaction cross sections
measured here and in other works\footnote{
MI85 -- Nucl. Phys. A {\bf 441} (1985) 617;
MI86 -- NIM B {\bf 16} (1986) 61;
MI89 -- Analyst {\bf 114} (1989) 287;
Mi90 -- NEANDC(E)-312-U(1990) 46;
MI93 -- INDC(GER)-037/LN(1993) 49;
MI95 -- NIM B {\bf 103} (1995) 183;
MI96 -- NIM B {\bf 114} (1996) 91;
MI97 -- NIM B {\bf 129} (1997) 153.}.
The height-energy dependence 
of the detection efficiency
is displayed in Fig. \ref{fig4}.

The discrepancies between the two sets of high-energy data in 
Fig. \ref{fig3} have yet to be studied.
The analytical expression of the spectrometer detection efficiency 
as a function of energy and 
sample position height is
 \begin{eqnarray}
\label{other}
\varepsilon \left( E, H \right) = \varepsilon_{base} 
\left( E \right) \cdot \!\!\left[\frac{\left(q_1+ q_2\cdot 
\ln E + H_{base} \right)}
{\left( q_1 + q_2\cdot \ln E + H \right)} \right]^2 ,
\end{eqnarray}
where $q_1$ and $q_2$  are parameters defined by fitting the experimental 
results.
An analysis of the $\gamma$-spectrum processing codes has shown 
that the GENIE-2000 code is superior to 
the others because of its interactive mode of fitting the peaks, 
which permits correction of the 
automated computer-aided processing; therefore, we chose it for our work.

\section*{Experimental results and measurement errors}

Tables \ref{tab2} and \ref{tab3} present the results of measuring 
the reaction product yields in the 1 GeV proton-irradiated 
\nuc{208}{Pb} and 2.6 GeV proton-irradiated \nuc{nat}{W} samples. 
113 yields from \nuc{208}{Pb} have been obtained, 
of which, 6 independent yields (i), 17 independent yields of metastable 
states (m), 15 independent 
yields of metastable and ground states ($\Sigma m_j$+g), 
64 cumulative yields (c), and 11 supra cumulative 
yields, when the addend may exceed the determination error (c$^*$). 
107 yields from \nuc{nat}{W} are presented, 
of which, 6 independent yields (i), 9 metastable state yields (m), 
5 yields of metastable and ground 
states ($\Sigma m_j$+g), 86 cumulative yields (c), and 1 supra 
cumulative yield (c$^*$).

From Tables \ref{tab2} and \ref{tab3} one can see that the experimental 
errors are ranging within $\sim$(6-35)\%. The 
main contribution to the total error is from uncertainties in the 
nuclear data, namely, in the absolute 
quantum yields and cross sections of the monitor reactions.

\section*{Comparison with experimental data obtained elsewhere}
\noindent

Table \ref{tab1} and Fig. \ref{fig9} compare some of the present 
results with experimental data of other 
laboratories published in \cite{bib6}.

\begin{table}[h]  

\caption{Experimental product nuclide yields in 1 GeV 
proton-irradiated \nuc{208}{Pb} } \vspace{2pt}   \label{tab2}

\begin{center}
%\vspace*{0.5cm}
{\small
\begin{tabular}{|c|c|c|c|}\hline
{Product} & T$_{1/2}$ & Type & Yield (mb) \\ \hline
\nuc{206}{Bi}&  6.243d  &      i     &      4.60$\pm$     0.29\\
\nuc{205}{Bi}& 15.31d  &      i     &      6.20$\pm$     0.40\\
\nuc{204}{Bi}& 11.22h  &      i(m1+m2+g)     &      5.29$\pm$     0.80\\
\nuc{203}{Bi}& 11.76h  &      i(m+g)     &      4.84$\pm$     0.59\\
\nuc{204m}{Pb}& 67.2m  &      i(m)   &      11.0$\pm$      1.0\\
\nuc{203}{Pb}& 51.873h  &      c     &      31.5$\pm$      2.1\\
\nuc{201}{Pb}&  9.33h  &  c$^{*}$         &      26.9$\pm$      2.4\\
\nuc{200}{Pb}& 21.5h  &  c   &      18.2$\pm$      1.2\\
\nuc{198}{Pb}&  2.4h  &  c       &       8.9$\pm$      2.1\\
\nuc{197m}{Pb}& 43.0m  &  c$^{*}$            &      17.9$\pm$      4.0\\
\nuc{202}{Tl}& 12.23d  &  c       &      18.9$\pm$      1.2\\ \hline
   \end{tabular} 
}\end{center}
\end{table}

{\small
\begin{center}
  \begin{tabular}{|c|c|c|c|}
\multicolumn{4}{l}{Continuation of Table 1}
\\ \hline
{Product} & T$_{1/2}$ & Type & Yield (mb) \\ 
\hline
\nuc{201}{Tl}& 72.912h  &  c           &      43.7$\pm$      2.9\\
\nuc{200}{Tl}& 26.1h  &  c         &      40.6$\pm$      2.6\\
\nuc{200}{Tl}& 26.1h  &  i(m+g)         &      22.7$\pm$      1.5\\
\nuc{199}{Tl}&  7.42h  &  c         &      38.5$\pm$      5.2\\
\nuc{198m1}{Tl}&  1.87h  &  i(m1+m2)     &      17.6$\pm$      3.6\\
\nuc{198}{Tl}&  5.30h  &  c         &      35.9$\pm$      5.0\\
\nuc{196m}{Tl}& 84.6m  &  i(m)         &      34.8$\pm$      4.4\\
\nuc{203}{Hg}& 46.612d  &  c         &      4.03$\pm$     0.27\\
\nuc{197m}{Hg}& 23.8h  &  i(m)         &      10.7$\pm$      0.7\\
\nuc{195m}{Hg}& 41.6h  &  i(m)         &      13.6$\pm$      2.0\\
\nuc{193m}{Hg}& 11.8h  &  i(m)         &      18.9$\pm$      2.5\\
\nuc{192}{Hg}&  4.85h  &  c         &      35.2$\pm$      2.8\\
\nuc{198m }{Au}& 54.48h  &  i(m)           &      1.01$\pm$     0.14\\
\nuc{198}{Au}& 64.684h  &  i(m+g)         &      2.11$\pm$     0.22\\
\nuc{198}{Au}& 64.684h  &  i           &      1.09$\pm$     0.30\\ 
\nuc{196}{Au}&  6.183d  &  i(m1+m2+g)  &      4.13$\pm$     0.35\\
\nuc{195}{Au}& 186.098d &  c           &      48.7$\pm$      5.5\\
\nuc{194}{Au}& 38.020h  &  i(m1+m2+g) &      7.06$\pm$     0.75\\
\nuc{192}{Au}&  4.94h  &  c           &      46.9$\pm$      6.6\\
\nuc{192}{Au}&  4.94h  &  i(m1+m2+g) &      11.6$\pm$      1.7\\
\nuc{191}{Pt}& 69.6h  &  c     &      40.1$\pm$      4.4\\
\nuc{189}{Pt}& 10.87h  &  c         &      46.8$\pm$      4.8\\
\nuc{188}{Pt}& 10.2d  &  c     &      40.5$\pm$      2.9\\
\nuc{190}{Ir}& 11.78d  &  c         &      0.69$\pm$     0.06\\
\nuc{188}{Ir}& 41.5h  &  c     &      43.2$\pm$      3.2\\
\nuc{188}{Ir}& 41.5h  &   i(m+g)  &      2.93$\pm$     0.69\\
\nuc{186}{Ir }& 16.64h  &  c$^{*}$         &      20.8$\pm$      1.9\\
\nuc{185}{Ir}& 14.4h  &  c$^{*}$         &      34.8$\pm$      2.3\\
\nuc{184}{Ir}&  3.09h  &  c$^{*}$         &      39.5$\pm$      3.0\\
\nuc{185}{Os}& 93.6d  &  c         &      41.8$\pm$      2.8\\
\nuc{183m}{Os}&  9.9h  &  i(m)         &      23.2$\pm$      1.5\\
\nuc{182}{Os}& 22.1h  &  c         &      42.0$\pm$      2.8\\
\nuc{183}{Re}& 70.0d  &  c         &      41.7$\pm$      2.9\\
\nuc{182}{Re}& 12.7h  &  c         &      45.2$\pm$      3.7\\
\nuc{181}{Re}& 19.9h  &  c         &      43.1$\pm$      5.9\\
\nuc{179}{Re}& 19.7m  &  c$^{*}$&      47.8$\pm$      4.2\\
\nuc{177}{W}&  2.25h  &  c         &      30.1$\pm$      3.5\\
\nuc{176}{W}&  2.30h  &  c         &      30.8$\pm$      4.3\\
\nuc{176}{Ta}&  8.09h  &  c           &      34.5$\pm$      3.6\\ 
\nuc{173}{Ta}&  3.14h  &  c         &      31.0$\pm$      3.9\\
\nuc{172}{Ta}& 36.8m  &  c$^{*}$&      17.3$\pm$      2.3\\
\nuc{175}{Hf}& 70.0d  &  c         &      31.3$\pm$      2.3\\
\nuc{173}{Hf}& 23.6h  &  c         &      28.4$\pm$      2.6\\
\nuc{172}{Hf}& 683.017d &  c         &      24.1$\pm$      1.6\\
\nuc{171}{Hf}& 12.1h  &  c         &      18.2$\pm$      2.8\\
\nuc{170}{Hf}& 16.01h  &  c         &      22.1$\pm$      6.8\\
\nuc{172}{Lu}&  6.7d  &  c         &      23.9$\pm$      1.7\\
\nuc{172}{Lu}&  6.7d  &  i(m+g)  &      0.19$\pm$     0.05\\
\nuc{171}{Lu}&  8.24d  &  c         &      26.1$\pm$      1.8\\
\nuc{170}{Lu}& 48.288h  &  c         &      21.7$\pm$      2.9\\
  \hline  \end{tabular} 
 \end{center} }

%\begin{table} 
%\label{results}
%\caption{continuation of table } 
%Continuation of table \ref{results}.\vspace{8pt}   \\

 {\small
\begin{center}
  \begin{tabular}{|c|c|c|c|}
\multicolumn{4}{l}{Continuation of Table 1} \\ \hline
{Product} & T$_{1/2}$ & Type & Yield (mb) \\ 
\hline
\nuc{169}{Lu}& 34.06h  &  c         &      18.6$\pm$      1.2\\
\nuc{169}{Yb}& 32.026d  &  c         &      20.9$\pm$      1.5\\
\nuc{166}{Yb}& 56.7h  &  c         &      16.1$\pm$      1.1\\
\nuc{167}{Tm}&  9.25d  &  c         &      19.4$\pm$      4.0\\
\nuc{165}{Tm}& 30.06h  &  c         &      14.4$\pm$      1.4\\
\nuc{160}{Er}& 28.58h  &  c         &      8.8$\pm$      0.6\\
\nuc{157}{Dy}&  8.14h  &  c         &      5.73$\pm$     0.45\\
\nuc{155}{Dy}&  9.90h  &  c$^{*}$ &      3.66$\pm$     0.27\\
\nuc{155}{Tb}&  5.32d  &  c         &      4.16$\pm$     0.39\\
\nuc{153}{Tb}& 56.16h  &  c$^{*}$ &      2.52$\pm$     0.25\\
\nuc{152}{Tb}& 17.50h  &  c$^{*}$&      2.10$\pm$     0.17\\
\nuc{153}{Gd}& 241.6d &  c         &      2.60$\pm$     0.23\\
\nuc{149}{Gd}&  9.28d  &  c         &      2.24$\pm$     0.18\\
\nuc{146}{Gd}& 48.27d  &  c         &      1.26$\pm$     0.09\\ 
\nuc{147}{Eu}& 24.0d  &  c         &      0.98$\pm$     0.31\\
\nuc{146}{Eu}&  4.59d  &  c         &      1.63$\pm$     0.11\\
\nuc{146}{Eu}&  4.59d  &  i         &      0.37$\pm$     0.05\\
\nuc{143}{Pm}& 265.0d &  c         &      1.02$\pm$     0.13\\
\nuc{139}{Ce}& 137.64d &  c         &      0.83$\pm$     0.06\\
\nuc{121m}{Te}& 154.0d &  i(m)    &      0.44$\pm$     0.04\\
\nuc{121}{Te}& 16.78d  &  c         &      1.11$\pm$     0.11\\
\nuc{119m}{Te}&  4.7d  &  i(m)        &      0.40$\pm$     0.04\\
\nuc{120m}{Sb}&  5.76d  &  i(m)       &      0.54$\pm$     0.05\\
\nuc{114m}{In}& 49.51d  &  i(m1+m2) &      0.95$\pm$     0.19\\
\nuc{110m}{Ag}& 249.79d &  i(m)     &      1.12$\pm$     0.09\\
\nuc{106m}{Ag}&  8.28d  &  i(m)           &      0.89$\pm$     0.08\\
\nuc{105}{Ag}& 41.29d  &  c         &      0.65$\pm$     0.12\\
\nuc{105}{Rh}& 35.36h  &  c           &      4.63$\pm$     0.54\\
\nuc{101m}{Rh}&  4.34d  &  i(m)           &      1.29$\pm$     0.16\\
\nuc{103}{Ru}& 39.26d  &  c         &      3.84$\pm$     0.26\\
\nuc{96}{Tc}&  4.28d  &  i(m+g)   &      1.20$\pm$     0.09\\
\nuc{95}{Tc}& 20.0h  &  c         &      1.38$\pm$     0.13\\
\nuc{96}{Nb}& 23.35h  &  i         &      2.31$\pm$     0.19\\
\nuc{95}{Nb}& 34.975d  &  c         &      5.41$\pm$     0.34\\
\nuc{95}{Nb}& 34.975d  &  i(m+g)     &      3.03$\pm$     0.20\\
\nuc{95}{Zr}& 64.02d  &  c         &      2.34$\pm$     0.15\\
\nuc{89}{Zr}& 78.41h  &  c         &      2.30$\pm$     0.16\\
\nuc{88}{Zr}& 83.4d  &  c           &      0.76$\pm$     0.08\\
\nuc{90m}{Y}&  3.19h  &  i(m)      &      4.82$\pm$     0.39\\
\nuc{88}{Y}& 106.65d &  c         &      4.03$\pm$     0.27\\
\nuc{88}{Y}& 106.650d &  i(m+g)  &      3.41$\pm$     0.25\\
\nuc{87}{Y}& 79.8h  &  c$^{*}$&      2.94$\pm$     0.23\\ 
\nuc{85}{Sr}& 64.84d  &  c         &      2.76 $\pm$     0.22\\
\nuc{86}{Rb}& 18.631d  &  i(m+g)  &      5.48 $\pm$     0.66\\
\nuc{83}{Rb}& 86.2d  &  c         &      3.46 $\pm$     0.28\\
\nuc{82m}{Rb}&  6.472h  &  i(m)       &      2.73 $\pm$     0.30\\
\nuc{82}{Br}& 35.3h  &  i(m+g)    &      2.17 $\pm$     0.14\\
\nuc{75}{Se}& 119.77d &  c         &      1.34 $\pm$     0.09\\
\nuc{74}{As}& 17.77d  &  i         &      1.86 $\pm$     0.18\\
\nuc{59}{Fe}& 44.503d  &  c         &      0.91 $\pm$     0.08\\
\nuc{65}{Zn}& 244.26d  &  c         &      0.79 $\pm$     0.19\\
\nuc{46}{Sc}& 83.81d  &  i(m+g)  &      0.35 $\pm$     0.06\\ \hline

   \end{tabular} 
\end{center} }
%\end{table}
 
\clearpage

\begin{table}  

%\vspace*{-0.5cm}
\caption{Experimental product nuclide yields in 
2.6 GeV proton-irradiated \nuc{nat}{W} } 
\vspace{2pt}   \label{tab3}

\begin{center}
{\small

\begin{tabular}{|c|c|c|c|}\hline
{Product} & T$_{1/2}$ & Type & Yield (mb) \\ \hline
\nuc{	177}{W	} &	2.25h	&	c	&	13.9	$\pm$	1.9	 \\
\nuc{	176}{W	} &	2.30h	&	c	&	9.9	$\pm$	2.9	 \\
\nuc{	184}{Ta	} &	8.7h	&	c	&	4.44	$\pm$	0.43	 \\
\nuc{	183}{Ta	} &	5.1d	&	c	&	10.5	$\pm$	1.0	 \\
\nuc{	182}{Ta	} &	114.43d	&	c	&	12.9	$\pm$	1.3	 \\
\nuc{	178m}{Ta	} &	2.36h	&	i(m)	&	8.1	$\pm$	1.3	 \\
\nuc{	176}{Ta	} &	8.09h	&	c	&	29.3	$\pm$	3.3	 \\
\nuc{	175}{Ta	} &	10.5h	&	c	&	26.0	$\pm$	2.8	 \\
\nuc{	174}{Ta	} &	63m	&	c	&	25.8	$\pm$	2.8	 \\
\nuc{	181}{Hf	} &	42.39	&	c	&	1.26	$\pm$	0.12	 \\
\nuc{	173}{Hf	} &	23.6h	&	c	&	29.9	$\pm$	2.5	 \\
\nuc{	171}{Hf	} &	12.1h	&	c	&	19.6	$\pm$	2.4	 \\
\nuc{	170}{Hf	} &	16.01h	&	c	&	19.6	$\pm$	4.0	 \\
\nuc{	172}{Lu	} &	6.7d	&	i(m+g)	&	4.32	$\pm$	0.56	 \\
\nuc{	171}{Lu	} &	8.24d	&	c	&	30.1	$\pm$	2.46	 \\
\nuc{	171}{Lu	} &	8.24d	&	i(m+g)	&	10.8	$\pm$	2.0	 \\
\nuc{	170}{Lu	} &	48.288h	&	c	&	24.8	$\pm$	2.2	 \\
\nuc{	169}{Lu	} &	34.06h	&	c	&	22.2	$\pm$	1.8	 \\
\nuc{	167}{Lu	} &	51.5m	&	c	&	23.4	$\pm$	2.4	 \\
\nuc{	167}{Yb	} &	17.5m	&	c	&	24.9	$\pm$	2.8	 \\
\nuc{	166}{Yb	} &	56.7h	&	c	&	24.6	$\pm$	2.1	 \\
\nuc{	166}{Tm	} &	7.7h	&	c	&	27.2	$\pm$	2.3	 \\
\nuc{	166}{Tm	} &	7.7h	&	i	&	2.36	$\pm$	0.46	 \\
\nuc{	165}{Tm	} &	30.06h	&	c	&	27.1	$\pm$	2.4	 \\
\nuc{	163}{Tm	} &	1.81h	&	c	&	26.3	$\pm$	3.3	 \\
\nuc{	161}{Tm	} &	33m	&	c	&	21.0	$\pm$	2.5	 \\
\nuc{	161}{Er	} &	3.21h	&	c	&	24.3	$\pm$	2.5	 \\
\nuc{	160}{Er	} &	28.58h	&	c	&	23.9	$\pm$	2.2	 \\
\nuc{	157}{Er	} &	25m	&	c	&	26.4	$\pm$	5.9	 \\
\nuc{	156}{Er	} &	19.5m	&	c	&	15.9	$\pm$	2.4	 \\
\nuc{	160m1}{Ho	} &	5.02h	&	i(m1+m2)	&	24.9	$\pm$	2.3	 \\
\nuc{	157}{Ho	} &	12.6m	&	c	&	26.1	$\pm$	6.8	 \\
\nuc{	156}{Ho	} &	56m	&	c	&	19.6	$\pm$	1.8	 \\
\nuc{	157}{Dy	} &	8.14h	&	c	&	24.2	$\pm$	2.2	 \\
\nuc{	155}{Dy	} &	9.90h	&	c	&	22.1	$\pm$	1.9	 \\
\nuc{	153}{Dy	} &	6.4h	&	c	&	14.0	$\pm$	1.9	 \\
\nuc{	152}{Dy	} &	2.38h	&	c	&	15.6	$\pm$	1.3	 \\
\nuc{	155}{Tb	} &	5.32d	&	c	&	22.7	$\pm$	1.9	 \\
\nuc{	153}{Tb	} &	56.16h	&	c	&	18.9	$\pm$	1.7	 \\
\nuc{	152}{Tb	} &	17.50h	&	c	&	16.2	$\pm$	1.3	 \\
\nuc{	151}{Tb	} &	17.609h	&	c	&	16.7	$\pm$	1.4	 \\
\nuc{	149}{Tb	} &	4.118h	&	c	&	6.85	$\pm$	0.62	 \\
\nuc{	147}{Tb	} &	1.70h	&	c	&	2.15	$\pm$	0.34	 \\
\nuc{	151}{Gd	} &	124.0d	&	c	&	19.0	$\pm$	2.2	 \\
\nuc{	149}{Gd	} &	9.28d	&	c	&	20.4	$\pm$	1.7	 \\
\nuc{	147}{Gd	} &	38.1h	&	c	&	18.6	$\pm$	1.6	 \\
\nuc{	146}{Gd	} &	48.27d	&	c	&	19.4	$\pm$	1.6	 \\
\nuc{	145}{Gd	} &	23.0m	&	c	&	12.9	$\pm$	1.4	 \\
\nuc{	149}{Eu	} &	93.1d	&	c	&	26.7	$\pm$	3.4	 \\ \hline
\end{tabular} }
\end{center}
\end{table}

\clearpage

{\small
\begin{center}

\begin{tabular}{|c|c|c|c|}
\multicolumn{4}{l}{Continuation of Table 2} \\ \hline
{Product} & T$_{1/2}$ & Type & Yield (mb) \\ 
\hline
\nuc{	147}{Eu	} &	24.0d	&	c	&	22.4	$\pm$	2.0	 \\
\nuc{	146}{Eu	} &	4.59d	&	c	&	23.0	$\pm$	1.9	 \\
\nuc{	146}{Eu	} &	4.59d	&	i	&	3.62	$\pm$	0.31	 \\
\nuc{	145}{Eu	} &	5.93d	&	c	&	17.8	$\pm$	1.6	 \\
\nuc{	139}{Nd	} &	5.5h	&	c	&	2.87	$\pm$	0.43	 \\
\nuc{	139}{Ce	} &	137.64d	&	c	&	19.8	$\pm$	1.6	 \\
\nuc{	135}{Ce	} &	17.7h	&	c	&	17.8	$\pm$	1.5	 \\
\nuc{	132}{Ce	} &	3.51h	&	c	&	16.3	$\pm$	2.7	 \\
\nuc{	132}{La	} &	4.8h	&	c	&	14.5	$\pm$	1.6	 \\
\nuc{	131}{Ba	} &	11.50d	&	c	&	16.2	$\pm$	1.3	 \\
\nuc{	126}{Ba	} &	100m	&	c	&	7.9	$\pm$	1.1	 \\
\nuc{	129}{Cs	} &	32.06h	&	c	&	18.7	$\pm$	1.6	 \\
\nuc{	127}{Xe	} &	36.4d	&	c	&	15.4	$\pm$	1.3	 \\
\nuc{	125}{Xe	} &	16.9h	&	c	&	14.2	$\pm$	1.2	 \\
\nuc{	123}{Xe	} &	2.08h	&	c	&	15.6	$\pm$	1.3	 \\
\nuc{	122}{Xe	} &	20.1h	&	c	&	11.7	$\pm$	1.0	 \\
\nuc{	121}{Te	} &	16.78d	&	c	&	10.7	$\pm$	1.1	 \\
\nuc{	119}{Te	} &	16.03h	&	c	&	9.17	$\pm$	0.74	 \\
\nuc{	119m}{Te	} &	4.7d	&	i(m)	&	1.97	$\pm$	0.17	 \\
\nuc{	117}{Te	} &	62m	&	c	&	8.81	$\pm$	0.77	 \\
\nuc{	118m}{Sb	} &	5.0h	&	i(m)	&	1.08	$\pm$	0.22	 \\
\nuc{	115}{Sb	} &	32.1m	&	¤$^*$	&	9.85	$\pm$	0.88	 \\
\nuc{	113}{Sn	} &	115.09d	&	¤	&	7.55	$\pm$	0.67	 \\
\nuc{	111}{In	} &	2.8049d	&	¤	&	7.44	$\pm$	0.74	 \\
\nuc{	110m}{In	} &	4.9h	&	i(m)	&	3.29	$\pm$	0.29	 \\
\nuc{	109}{In	} &	4.2h	&	c	&	5.12	$\pm$	0.43	 \\
\nuc{	106m}{Ag	} &	8.28d	&	i(m)	&	1.70	$\pm$	0.16	 \\
\nuc{	105}{Ag	} &	41.29d	&	c	&	5.33	$\pm$	0.69	 \\
\nuc{	100}{Pd	} &	87.12h	&	c	&	1.24	$\pm$	0.27	 \\
\nuc{	100}{Rh	} &	20.8h	&	c	&	3.97	$\pm$	0.44	 \\
\nuc{	100}{Rh	} &	20.8h	&	i	&	2.68	$\pm$	0.28	 \\
\nuc{	99m}{Rh	} &	4.7h	&	c	&	2.41	$\pm$	0.28	 \\
\nuc{	97}{Ru	} &	69.6h	&	c	&	3.13	$\pm$	0.28	 \\
\nuc{	96}{Tc	} &	4.28d	&	i(m+g)	&	1.73	$\pm$	0.20	 \\
\nuc{	93m}{Mo	} &	6.85h	&	i(m)	&	1.61	$\pm$	0.13	 \\
\nuc{	90}{Nb	} &	14.6h	&	c	&	2.58	$\pm$	0.22	 \\
\nuc{	89}{Zr	} &	78.41h	&	c	&	3.46	$\pm$	0.28	 \\
\nuc{	88}{Zr	} &	83.4d	&	c	&	2.56	$\pm$	0.27	 \\
\nuc{	88}{Y	} &	106.65d	&	c	&	3.49	$\pm$	0.34	 \\
\nuc{	88}{Y	} &	106.65d	&	i(m+g)	&	1.56	$\pm$	0.22	 \\
\nuc{	87}{Y	} &	79.8h	&	c	&	4.13	$\pm$	0.34	 \\
\nuc{	83}{Sr	} &	32.41h	&	c	&	1.96	$\pm$	0.93	 \\
\nuc{	84}{Rb	} &	32.77d	&	i (m+g)	&	1.31	$\pm$	0.14	 \\
\nuc{	83}{Rb	} &	86.2d	&	c	&	3.34	$\pm$	0.58	 \\
\nuc{	82m}{Rb	} &	6.472h	&	i(m)	&	1.89	$\pm$	0.17	 \\
\nuc{	77}{Kr	} &	74.4m	&	c	&	1.71	$\pm$	0.18	 \\
\nuc{	75}{Se	} &	119.77d	&	c	&	2.38	$\pm$	0.22	 \\
\nuc{	73}{Se	} &	7.15h	&	c	&	1.03	$\pm$	0.11	 \\
\nuc{	74}{As	} &	17.77d	&	c	&	1.38	$\pm$	0.16	 \\
\nuc{	69m}{Zn	} &	13.76h	&	i(m)	&	0.42	$\pm$	0.038	 \\
\nuc{	54}{Mn	} &	312.12d	&	i	&	2.51	$\pm$	0.42	 \\
\nuc{	51}{Cr	} &	27.704d	&	c	&	4.5	$\pm$	1.4	 \\ \hline
\end{tabular} 
\end{center} }

{\small
\begin{center}

\begin{tabular}{|c|c|c|c|}
\multicolumn{4}{l}{Continuation of Table 2} \\ \hline
{Product} & T$_{1/2}$ & Type & Yield (mb) \\ 
\hline
\nuc{	48}{V	} &	15.973d	&	c	&	0.557	$\pm$	0.062	 \\
\nuc{	48}{Sc	} &	43.67h	&	i	&	0.668	$\pm$	0.091	 \\
\nuc{	43}{K	} &	22.3h	&	c	&	0.681	$\pm$	0.084	 \\
\nuc{	28}{Mg	} &	20.91h	&	c	&	0.91	$\pm$	0.089	 \\
\nuc{	24}{Na	} &	14.959h	&	c	&	4.09	$\pm$	0.34	 \\
\nuc{	7}{Be	} &	53.29d	&	i	&	8.7	$\pm$	1.0	 \\ \hline
\end{tabular} 
\end{center} }

\section*{Simulation of experimental results}
\noindent

Simulation techniques are of essential importance when forming 
the set of nuclear constants to be 
used in designing the ADS facilities, because they are universal 
and save much time and labour. At the 
same time, the present-day accuracy and reliability of the simulated 
results are inferior to experiment. 
Besides, the simulation codes are of different abilities to work when 
used to study the reactions that 
are of practical importance.

Considering the above, the present work is primarily aimed at verifying 
the simulation codes used 
most extensively for the above purpose with a view to not only estimating 
their ability to work when 
applied to the issues discussed here, but also opening up ways to 
improve them.

The following eight simulation codes were examined to meet these requirements:
\begin{itemize}
 \item  the CEM95 cascade-exciton code \cite{bib7},
 \item the CASCADE cascade-evaporation-fission-transport code \cite{bib8},
 \item the INUCL cascade-preequilibrium-evaporation-fission code \cite{bib9},
 \item the LAHET (ISABEL and Bertini options) 
 cascade-evaporation-fission code \cite{bib10},
 \item  the YIELDX semi-phenomenological code \cite{bib11},
 \item the CASCADE/INPE cascade-preequilibrium-evaporation-fission-transport 
 code \cite{bib12},
 \item the CEM2k cascade-exciton code \cite{cem2k}, a last 
 modification of the CEM95 code,
\end{itemize}

Contrary to the simulation results, the experimental data include not only 
the independent, but also 
(and mainly) cumulative and the supra cumulative yields
of residual product nuclei. To get a correct comparison 
between the experimental and simulation results, theoretical cumulative 
yields must be calculated on 
the basis of the simulated independent yields.

\begin{table}  
\begin{center}
\vspace*{-0.5cm}
\caption{The yields (mb) of some products in the 1 GeV proton-irradiated 
\nuc{208}{Pb} inferred from 
measurements at different laboratories;
the ZSR and GSI data are taken from [6], the ITEP data
are our present results} 
\vspace{2pt}   
\label{tab1}

\vspace*{0.5cm}
\begin{tabular}{|c|c|c|c|}\hline
Product nuclide	&	ZSR Hannover	&	ITEP	&	GSI Darmstadt	 \\ \hline
\nuc{200}{Tl}	&	22.3$\pm$ 6.1	&	22.7$\pm$ 1.5	&	17.0$\pm$ 0.4(1.6)	 \\ \hline
\nuc{196}{Au}	&	3.88$\pm$ 0.47	&	4.13$\pm$ 0.35	&	4.0$\pm$ 0.1(0.4)	 \\ \hline
\nuc{194}{Au}	&	6.85$\pm$ 0.92	&	7.06$\pm$ 0.75	&	6.3$\pm$ 0.2(0.6)	 \\ \hline
\nuc{148}{Eu}	&	0.104$\pm$ 0.04	&	--	&	0.075$\pm$ 0.005(0.010)	 \\ \hline
\nuc{144}{Pm}	&	0.068$\pm$ 0.013	&	--	&	0.036$\pm$ 0.003(0.006)	 \\ \hline
 \end{tabular} \end{center}
\end{table}

Since any branched isobaric chain can be presented to be a 
superposition of a few linear chains, 
the simulated cumulative and supra cumulative yields of a n-th 
nuclide can be calculated as
\begin{equation}\label{branch1}
\sigma^{cum}_n =  \sigma^{ind}_n + \sum^{n-1}_{i=1} 
\sigma^{ind}_i \prod^{n-1}_{j=i} \nu_j \mbox{,}
\end{equation}
\begin{equation}\label{branch2} \begin{array}{l}
\dsp \sigma_n^{cum^{\dsp*}} = \sigma_n^{ind} + 
\frac{\lambda_{n-1}}{\lambda_{n-1} - \lambda_{n}} 
\nu_{n-1}   \dsp
\times \left[\sigma_{n-1}^{ind} + \sum^{n-2}_{i=1} 
\left( \sigma_i^{ind} \prod^{n-2}_{j=i} \nu_j \right)\right] \end{array} 
\mbox{ .}
\end{equation}
The branching ratios of the decay chains were retrieved from \cite{bib13}.
To get a correct comparison between results by different codes, the 
calculations 
were renormalized to unified cross sections for proton-nucleus 
inelastic interactions from \cite{bib14}.

If an experiment-simulation difference of not 
above 30\% (0.77$<\sigma_{calc}/\sigma_{exp}<$1.3) 
is taken to be the coincidence criterion \cite{bib15}, 
the simulation accuracy can be presented to be the ratio of the number 
of such coincidences to the number of the comparison events. 
The 30\% level meets the accuracy requirements 
of the cross sections for nuclide production to be used in designing the 
ADS plants, according to \cite{bib15}. The mean 
simulated-to-experimental data ratio can be used as another coincidence 
criterion:
\begin{equation}
\label{coin}
<F> \; = 10^{\sqrt{\dsp <\log \left(\sigma_{cal, i}/\sigma_{exp, 
i}\right)^2>}} ,
\end{equation}
with its standard deviation
\begin{equation}\label{deviat}
S(<F>) = \; <\left(\log\left(\sigma_{cal, i}/\sigma_{exp, i}\right) - 
\log(<F>)\right)^2> ,
\end{equation}
where $<>$  designates averaging over all N$_S$ 
number of the experimental and simulated results used in a comparison.

The mean ratio $<F>$  together with its standard deviation $S(<F>)$  
defines the interval 
[$<F>:S(<F>)$ , $<F>\times S(<F>)$]  that covers about 2/3 of the 
simulation-to-experiment ratios.

The two criteria are considered sufficient 
to derive conclusions about the predictive power of
a given code.
The default options were committed to practical usage of the simulation codes.

\section*{Comparison of data with simulation results}
\noindent

\begin{figure} %\vspace{1cm} 
\begin{center}
\includegraphics[angle=-90, width=16cm]{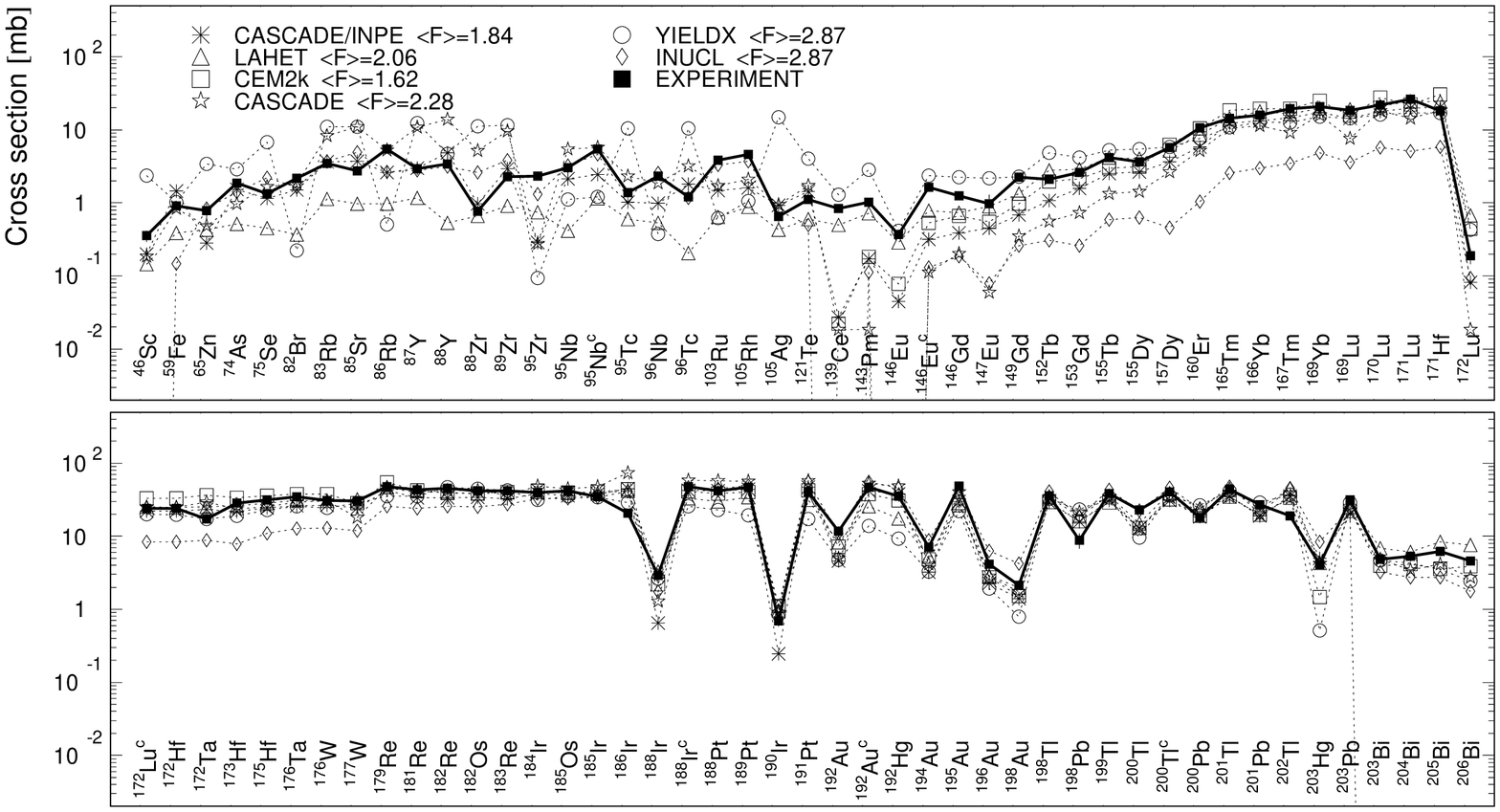}
\caption{Product comparison between the experimental (closed symbols) 
and simulated (open symbols) yields of radioactive reaction products 
from \nuc{208}{Pb} irradiated with 1 GeV protons. The 
cumulative yields are labeled with a ``c" 
when the respective independent yields are also shown.}
\label{fig5} 
\end{center} 
\end{figure}

\begin{figure} 
\vspace*{-0.8cm} 
\begin{center}
\includegraphics[angle=-90, width=16cm]{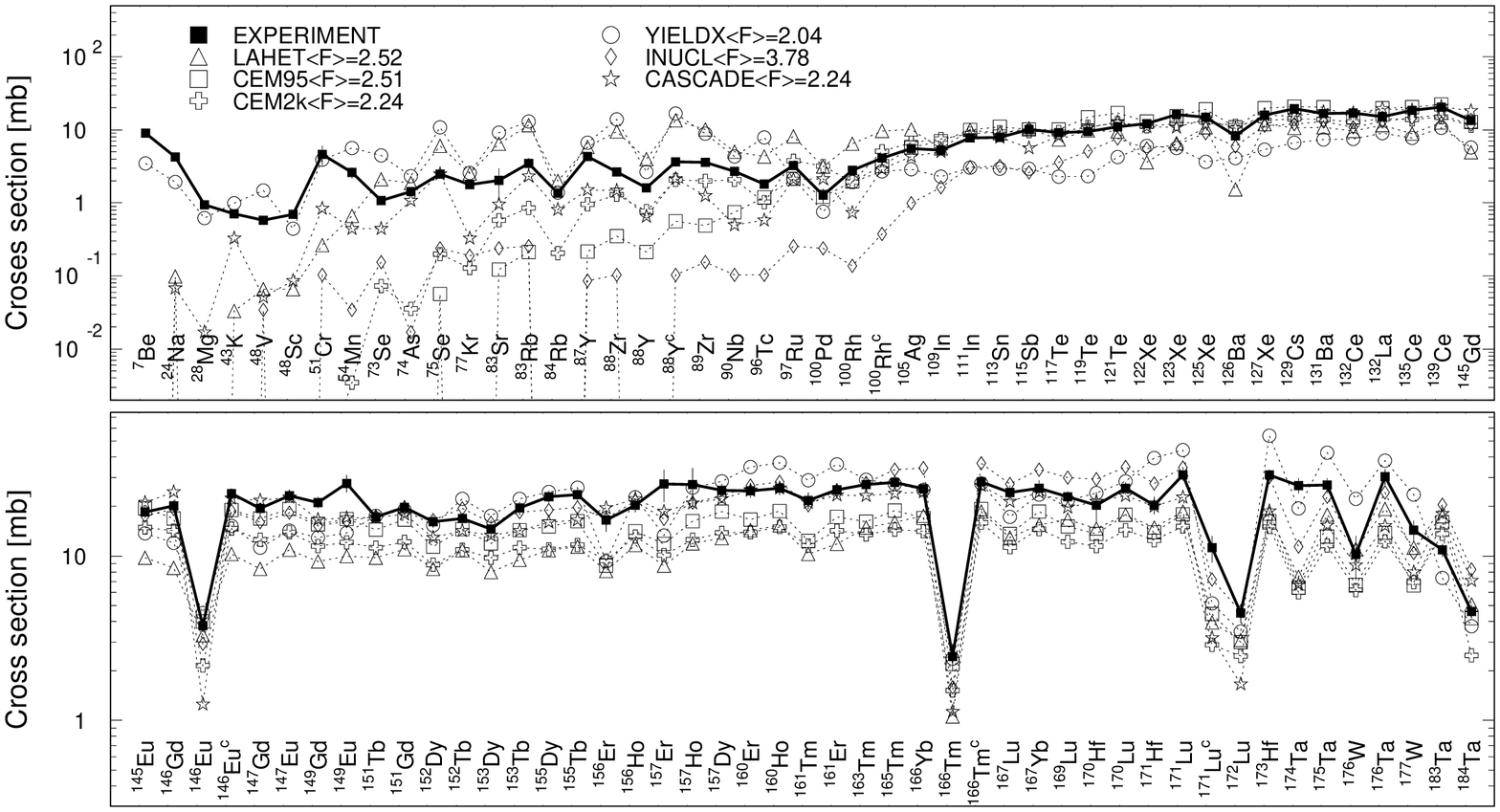}
\caption{Product comparison between the experimental (closed symbols) 
and simulated (open symbols) yields of radioactive reaction products 
from \nuc{nat}{W} irradiated with 2.6 GeV protons. The 
cumulative yields are labeled with a ``c" when the 
respective independent yields are also shown.}
\label{fig6} 
\end{center} 
\end{figure}

\begin{figure} %\vspace{1cm} 
\begin{center}
\includegraphics[angle=-90, width=15.2cm]{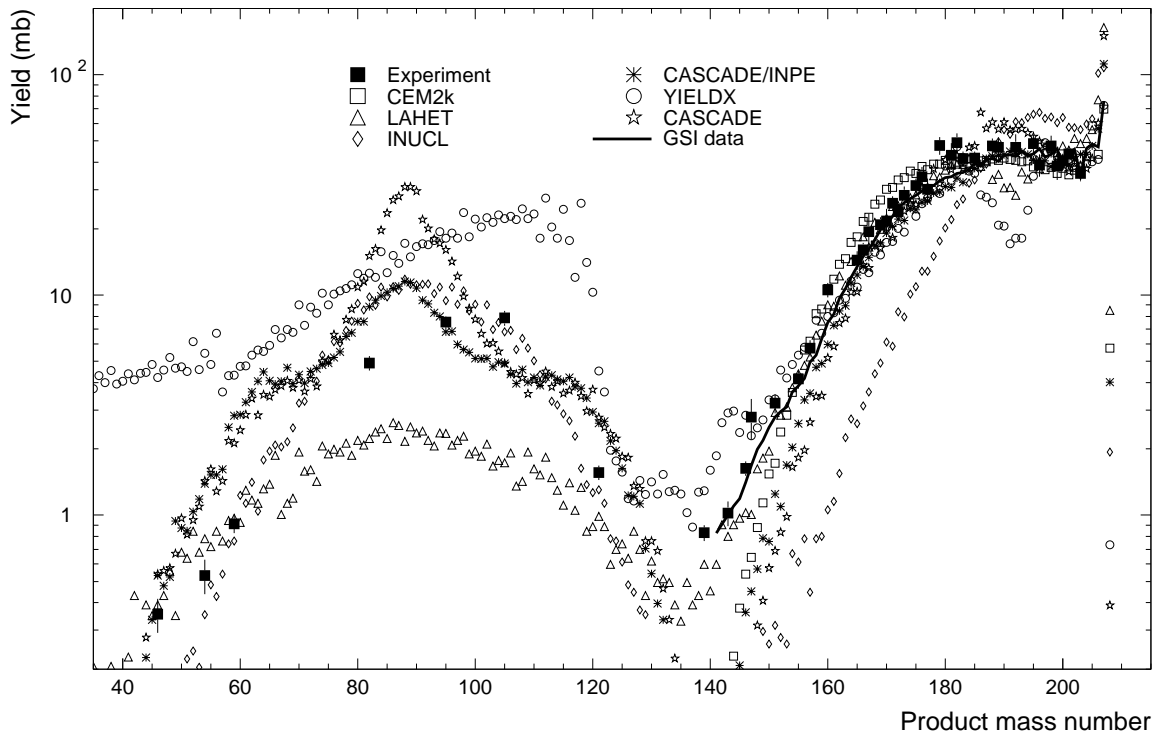}
\caption{Measured and calculated by the codes
mass product yields from \nuc{208}{Pb} irradiated with 1.0 GeV 
protons. 
For comparison, the GSI data from [6] are shown as well.}
\label{fig7} 
\end{center} 
\end{figure}

\begin{figure} %\vspace{1cm}
 \begin{center}
\includegraphics[angle=-90, width=15.2cm]{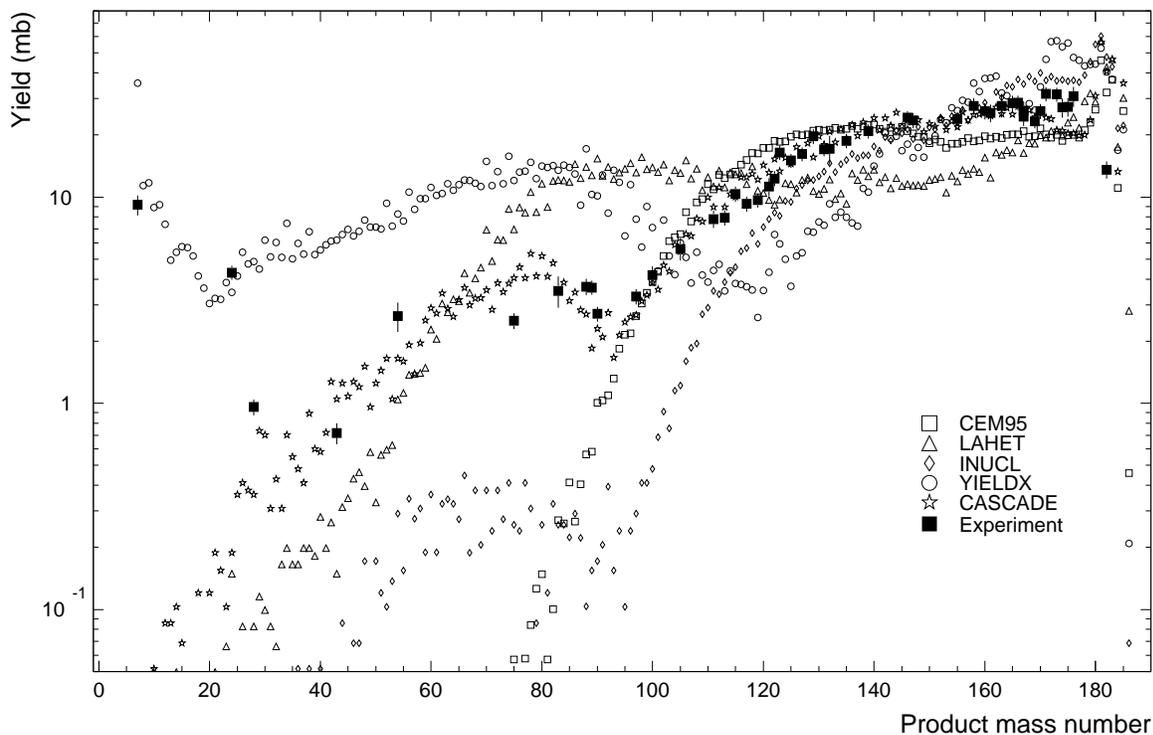}
\caption{Measured and calculated via the codes mass product yields 
from \nuc{nat}{W} irradiated with 2.6 GeV protons.}
\label{fig8} 
\end{center} 
\end{figure}

\begin{figure} %\vspace{1cm} 
\begin{center}
\includegraphics[angle=-90, width=15.2cm]{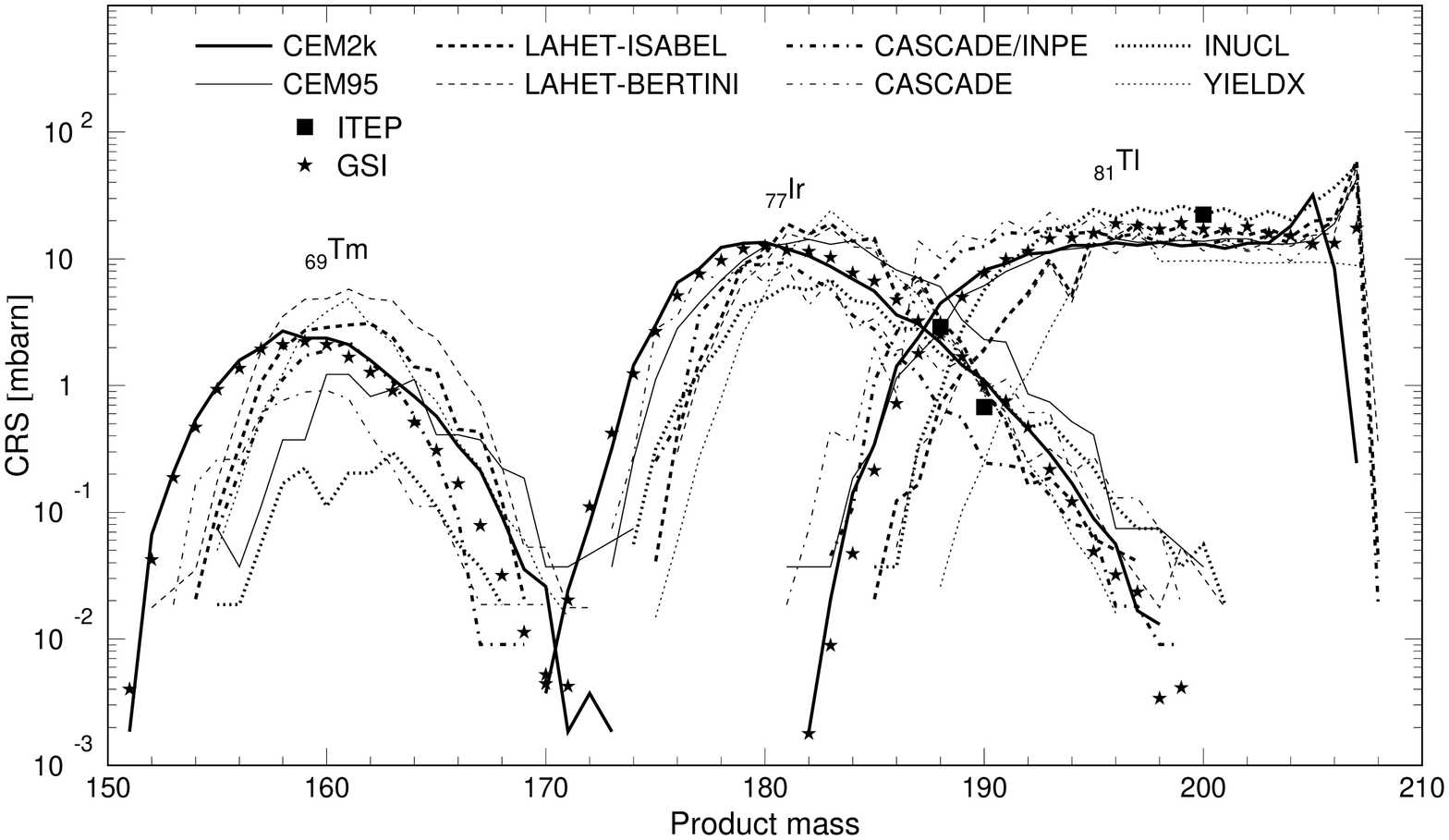}
\caption{Isotopic mass distributions of the reaction products in 
\nuc{208}{Pb}. The rich, inverse-kinematics, GSI data from [6] are 
shown by filled stars.}
\label{fig9} \end{center} \end{figure}

The results obtained with the above-mentioned codes are presented in:
\begin{itemize}
 \item Figs. \ref{fig5} and \ref{fig6},
  that show results of a detailed comparison between the simulated and 
experimental radioactive product yields;
 \item Figs. \ref{fig7} and \ref{fig8},
  that show the simulated mass distributions of reaction products together 
  with the measured cumulative (and supra cumulative) yields of the products 
  that are at an
immediate proximity to the stable isobar of a given mass 
(the sum of such yields from either 
sides in case both left- and right-hand branches of the chain are present). 
Obviously, the displayed simulation results do not contradict the 
experimental data if calculated values run 
above the experimental data and follow a general trend of the latter. This is 
because the 
direct $\gamma$-spectroscopy method used here identifies only 
radioactive products, that, as a rule, 
represents a 
significant fraction of the total mass yield, but, 
should a stable isobar of the given mass be 
produced, the
$\gamma$-spectroscopy data
are never equal to the total mass yield;
 \item In Fig. \ref{fig9}, 
 that shows the experimental and simulated independent yields of 
 reaction products in 
the form of isotopic mass distributions for several elements.
\end{itemize}

Table \ref{tab4} presents the statistics of our 
comparison between the experimental and simulated reaction 
product yields in the thin \nuc{208}{Pb} and \nuc{nat}{W} samples irradiated by 1.0 GeV and 2.6 GeV protons, 
respectively.
Namely, it shows
the total number of measured yields, N$_T$; the number of 
the measured yields selected 
to compare with calculations, N$_G$; 
the number of the product nuclei whose yields were simulated by 
a particular code, N$_S$; the number of the comparison events when the simulated data differ from the 
experimental results by not above 30\%, N$_{C1.3}$; the number of the comparison events when the 
simulated data differ from the experimental results by not more than a factor of 2.0, N$_{C2.0}$.

\begin{table} 
\caption{Comparison statistics for \nuc{208}{Pb} and \nuc{nat}{W} } \vspace{8pt}   
\label{tab4}
\begin{center}
\begin{tabular}{|c|c|c|c|c|c|c|}\hline
        & \multicolumn{3}{|c|}{Pb, E$_p$ = 1.0 GeV} & \multicolumn{3}{|c|}{W, E$_p$ = 2.6 GeV} \\
Code & \multicolumn{3}{|c|}{N$_T$ = 116, N$_G$ = 95} & \multicolumn{3}{|c|}{N$_T$ = 107, N$_G$ = 93} \\ \cline{2-7}
        & N$_{C_{1.3}}$/N$_{C_{2.0}}$/ &$<F>$&$S(<F>)$& N$_{C_{1.3}}$/N$_{C_{2.0}}$/ &$<F>$&$S(<F>)$\\
        & /N$_{S}$& &&/N$_{S}$ & &\\ \hline
LAHET	&	41/65/90	&	2.06	&	1.78	&	13/51/90	&	2.52	&	1.83 \\ \hline
CEM95	&	--	&	--	&	--	&	28/66/81	&	2.51	&	2.23 \\ \hline
CEM2k	&	38/58/66	&	1.62	&	1.44	&	17/60/84	&	2.24	&	1.73 \\ \hline
CASCADE&	33/60/86	&	2.28	&	1.90	&	48/71/91	&	2.24	&	2.04 \\ \hline
CASCADE/INPE&36/66/84	&	1.84	&	1.56	&	--		&	--	&	--      \\ \hline
INUCL	&	29/54/90	&	2.87	&	2.16	&	38/58/86	&	3.78	&	3.23 \\ \hline
YIELDX	&	30/54/90	&	2.87	&	2.24	&	25/60/93	&	2.04	&	1.58 \\ \hline
 \end{tabular} 
\end{center}
\end{table}

Since about 30\% of all measured
secondary nuclei are not spallation reaction products, 
an important criterion 
of the codes is their ability to simulate the high-energy fission and 
fragmentation processes.
Among the  codes used here, LAHET, CASCADE, INUCL, CASCADE/INPE, 
and YIELDX simulate both spallation and fission. 
The CEM95 and CEM2k codes simulate 
spallation only, which is explicitly reflected in a smaller 
number of the products simulated (the 
parameter N$_S$ in Table 4
and in the shapes of the simulation curves in Figs. 5-8.

The following conclusions follow from our analysis of the 
experiment-to-simulation comparison 
results presented in Table 4 and in Figs. \ref{fig5}-\ref{fig9}:
\begin{enumerate}
 \item Generally, all codes can quite adequately simulate the weak 
 spallation reactions (the A$\ge$180 
products for \nuc{208}{Pb} and the A$\ge$150 products for \nuc{nat}{W}), 
with the simulation results differing from 
experimental data within a factor of 2.
 \item In the deep spallation region (150$<$A$<$180 for \nuc{208}{Pb} 
 and 110$<$A$<$150 for \nuc{nat}{W}),
 the simulation codes are of very different predictive powers, namely,
\begin{itemize}
 \item the LAHET (when not shown explicitly as ``Bertini",
 all results by LAHET are of the ISABEL option), 
 CEM2k, CASCADE/INPE, and YIELDX predictions are 
 actually the same as 
the experimental data;
 \item the CASCADE code simulates the A$>$160 product yields adequately. 
 Below A = 160, however, the simulated data get underestimated progressively 
 (up to a factor of 5) compared with experiment (see Fig. 7);
 \item the INUCL code underestimates the yields of all the products 
 by a factor of 2-10 in all the above mass ranges (see Figs. 7 and 8).
\end{itemize}
 \item 
In the mass range characteristic of the fission products 
(50$<$A$<$150 for \nuc{208}{Pb} and 30$<$A$<$110 for 
\nuc{nat}{W}), the INUCL code predictions are in the best agreement 
with experiment when describing 
the yields from \nuc{208}{Pb}. As a rule, the INUCL-simulated results
differ from the data   by not 
above a factor of 1.5. In the case of \nuc{nat}{W}, however, the prediction 
quality deteriorates 
substantially. 
The LAHET-simulated yields are underestimated by a 
factor of 1.5-10.0 for Pb (Figs. 5 and 7) in the whole fission product
mass region and for A$<$60 in the case of W (Figs. 6 and 8)
but are oversetimated several times for fission fragments with
A$>$60 from W. 
The YIELDX-simulated yields are either under- or over-estimated by a 
factor of up to 30 without showing any physical regularities. 
The CASCADE/INPE-simulated yields of the 130$<$A$<$150 reaction products 
are strongly underestimated (up to 1-2 orders of magnitude), while the 
simulated 40$<$A$<$130 product yields agree with the data within
a factor of 2, as a rule. 
Generally, all the codes exhibit the feature noted above for INUCL, namely, 
the yield prediction quality in the case of \nuc{nat}{W} is much 
worse compared with \nuc{208}{Pb}, probably, because  the fission 
cross sections of high-excited compound nuclei with very low fissility 
are difficult to calculate.
 \item 
The last version of the improved cascade-exciton model code, CEM2k 
\cite{cem2k},
shows the best agreement with the 1 GeV Pb-data in the spallation region, 
especially for the isotopic mass distributions (Fig. 9).         
At 2.6 GeV (W-target), it overestimates the expected
experimental fission cross section of about 41 mb \cite{prokofiev00}
by a factor of 6.
This overestimation of the fission cross section causes an underestimation
of the yield of nuclei which are most likely to fission (with a very
low fissility) at the evaporation
stage of a reaction, after the cascade and preequilibrium stages , i.e.,
for 147 $<$ A $<$ 175 (see Fig. 6). 
Similar disagreement with the 2.6 GeV W-data 
one can see as
well for LAHET and CEM95, that is also related with an overestimation
of the fission cross section at 2.6 GeV (see Figs. 6 and 8). 
The code CEM2k is still under
development, its problem with the overestimation of fission cross
sections at energies above 1 GeV has yet to be solved, and it has to be 
complemented with a model of fission fragment production,
to be able to describe as well fission products.
\end{enumerate}

\section*{Conclusion}
\noindent

The trends shown by the advances in the nuclear transmutation 
of radioactive wastes and Spallation Neutron Source (SNS) facilities
permit us to 
expect that the accumulation and analysis study of nuclear data for ADS 
facilities will have the same 
rise of academic interest and practical commitments as in the nuclear 
reactor data during the last five 
decades. Therefore, the experimental data on the yields of the 
proton-induced reaction products as 
applied to the ADS  and SNS main targets and structure materials 
are urgent to accumulate. It should be 
emphasized  that the charge distributions in the isobaric 
decay chains are important to study as well. The 
data thus obtained would make it possible, first, to raise the 
information content of the comparisons between the experimental and 
simulated results and, second, to lift the uncertainties in 
experimental determination of the cumulative yields by establishing 
unambiguous relations between 
$\sigma^{cum}$ and  $\sigma^{cum^*}$ for many of the reaction product masses.

Regarding the codes benchmarked here, one may conclude that none
of them agree well with the data in the whole mass region
of product nuclides and all should be improved further.
The new CEM2k code developed recently at Los Alamos
\cite{cem2k} agrees  with our data
in the spallation region the best of the codes tested.
But CEM2k has yet to be completed by a model of fission fragmentation,
to become applicable in the fission-product region as well.

\section*{Acknowledgement}
\noindent

The authors are grateful to Prof. Vladimir Artisyuk (Tokyo Institute of Technology) for 
consulting and discussion our 
results on simulation-to-experimental comparisons.

The work was carried out under the ISTC Project \#839 supported by 
the European Community, Japan (JAERI), 
Norway and, partially, by the U. S. Department of Energy.

\end{document}